# BB84 Quantum Key Distribution System based on Silica-Based Planar Lightwave Circuits

[*]Yoshihiro NAMBU*, Takaaki HATANAKA,[†] and Kazuo NAKAMURA

(*) Corresponding author: E-mail address: y-nambu@ah.jp.nec.com

Fundamental and Environmental Research Laboratories, NEC, 34 Miyukigaoka, Tsukuba, Ibaraki 305-8501, Japan

[†]Fiber Optic Devices Division, NEC Corp., 747, Magi, Ohtsukimachi, Ohtsuki, Yamanashi 401-0016, Japan

We constructed an optical interferometer for a Bennett-Brassard quantum key distribution system using integrated optics based on planar lightwave circuit technology, and tested its operation and stability. Experimental results show that this interferometer is useful in implementing a practical quantum key distribution system.

Quantum key distribution (QKD) allows two remote parties, e.g., Alice and Bob,

to generate a secret key, with privacy ensured by quantum mechanics [1-2]. Since Bennett and Brassard's invented [1] and demonstrated [2] the BB84 protocol, numerous QKD systems that use optical-fiber-based techniques and faint laser pulses have been developed [3-11]. Recently, an alternative approach to the fabrication of a practical QKD system using integrated-optic interferometers instead of fiber-optic ones has been reported [12-14]. We showed that this integrated-optic interferometer is sufficiently stable and has potential to achieve longer transmission distances than the previous system. We demonstrated single-photon interference over 150 km [14]. Honjo et al. reported a QKD system using an integrated-optic interferometer based on the Bennett 92 protocol [15] for the first time [13].

In this letter, we describe one approach to the construction of a BB84 QKD system using an integrated-optic interferometer and faint laser pulses. We have developed a new hybrid device for the four-state preparation required by the BB84 protocol, which was fabricated by splicing together an integrated-optic phase modulator with a $LiNbO_3$ waveguiding structure and an asymmetric Mach-Zehnder interferometer (AMZ) on a silica-based planar lightwave circuit (PLC) platform. We demonstrated that the system operated as expected. Experimental results showed that our approach is promising for implementing a BB84 QKD system using an integrated-optic interferometer.

Figure 1 illustrates the setup. Alice's apparatus consists of a pulsed light source (LS), a symmetric Mach-Zehnder interferometer (SMZ) containing a phase modulator (PM) for fast data encoding in one of its arms, an AMZ, and an attenuator (ATT), while Bob's apparatus consists of an AMZ and two avalanche photodiodes (APDs). These two apparatuses are connected by single-mode optical fibers forming a complete QKD system. Details of the two AMZs are given by Kimura et al. [14]. The AMZs had the same 5 ns delay in one of their arms, and were fabricated on a silica-based PLC platform. Their optical loss was approximately 2 dB (excluding the 3 dB intrinsic loss at the coupler) and they were thermally controlled up to a precision of 0.01°C. Kimura et al. showed that optical interference, which was sufficiently stable and insensitive to polarization disturbance in the fiber link, was achieved by connecting the two AMZs with an optical fiber. The intrinsic extinction ratio for the observed interference was more than 20 dB.

We used a commercially available $LiNbO_3$ intensity modulator to form the SMZ section of Alice's apparatus. This modulator had two parallel $LiNbO_3$ waveguides, both ends of which were connected to Y-branches. An electrode was fabricated on one of the two waveguides. We cut one of the Y-branches, and spliced the end of the remaining two waveguides to the two ports of the 3 dB coupler of Alice's AMZ. This SMZ section acts as a variable ratio coupler, the branching ratio of which can be controlled by adjusting the voltage applied to the electrode. Alice can use it to prepare one of four states, namely, $|l\rangle$, $|s\rangle$, $|l\rangle+|s\rangle$, or $|l\rangle-|s\rangle$, on the optical fibre link, where $|l\rangle$ and $|s\rangle$ denote a single-photon state that had traveled via the short or long path of Alice's AMZ.

The function of Bob's apparatus is similar to that given by Tittel et al. [16], who reported a QKD system using entangled photons in energy-time Bell states. After photons travel through Bob's AMZ, Bob can find them in one of three time slots. The first slot corresponds to photons taking the short path in both AMZs, while the last slot corresponds to photons taking both the long paths, whereas the central one corresponds to photons taking the short path in Alice's AMZ and the long one in Bob's, and vice versa. Bob can discriminate either between $|l\rangle$ and $|s\rangle$ or between $|l\rangle+|s\rangle$ and $|l\rangle-|s\rangle$ if he was given information about the basis, that is, whether the given state was in the set $\{|l\rangle, |s\rangle\}$ or $\{|l\rangle+|s\rangle, |l\rangle-|s\rangle\}$. If the basis was $\{|l\rangle, |s\rangle\}$, Alice's choice between $|s\rangle$ and $|l\rangle$ correlates with photon-detection events in the first and last time slots, respectively. Bob can definitely identify the state that Alice sent for those events in which he observed photons in these time slots, depending on the time slot in which he observed photons. Those events in which photons were detected in the last and first time slots, therefore, correspond to measurements compatible with the basis $\{|l\rangle, |s\rangle\}$. If the basis was $\{|l\rangle+|s\rangle, |l\rangle-|s\rangle\}$, by contrast, the two possible paths for the photons arriving in the central slot interfere with each other. Alice's choice then correlates with the output ports of the photons at Bob's AMZ in the central slot. Bob can definitely identify the state that Alice sent for those events in which he observed photons in this time slot, depending on the port in which he observed them. Such events in which photons were detected in the central slot correspond to measurements compatible with the basis, $\{|l\rangle+|s\rangle, |l\rangle-|s\rangle\}$. In addition, there is no way to perfectly discriminate between the four states unless information on the basis is given, thereby enabling us to implement the BB84 QKD system.

There are several advantages in this system over conventional systems [3-5]. First, the phase-modulator in Bob's apparatus can be avoided, thereby reducing photon loss and error rate. Second, the key-generation rate increases twofold because the detection events in the last and first time slots, which are discarded in conventional systems, contribute to the generation of keys. Third, no random-number generator is necessary for choosing a measurement basis in Bob's apparatus because it is chosen by Nature [16]. This would improve security and reduce the cost of Bob's apparatus. Fourth, since the measurement basis is unknown even to Alice and Bob, a Trojan horse attack on Bob's apparatus is useless. It should be noted that there are costs to pay for these advantages. First, to implement a BB84 QKD system, we need to develop a more sophisticated photon-detection apparatus that can discern at which time slot, as well as at which port, the photons arrive. Second, to enable photon detection in the three time slots, the detector must be triggered three times more often than that in conventional systems, which increases the error rate due to the dark count in detectors. Although part of this increase in noise could be compensated for by the twofold increase in key-generation rate, this compensation is not complete. The readers may also think that the after-pulsing effect of APDs is problematic since the interval between adjacent time slots is small (5 ns). The problem can be solved, however, if we design the photon-detection apparatus to register only the time slot in which the APDs fired first.

To confirm the operation and the stability of our system, we used intense laser pulses $\lambda$ of 1.55 $\mu$m, ~ 1 ns long, with 1 MHz repetition and observed the output signal from Bob's AMZ with fast InGaAs PIN photo detectors and an oscilloscope. Alice's and Bob's apparatuses were connected by a short single-mode fiber, and a polarization scrambler was inserted to simulate random polarization fluctuation in the fiber. We adjusted the temperature of the two AMZs and the voltage applied to the LiNbO$_3$ modulator in such a way that the $\{|l\rangle, |s\rangle, |l\rangle+|s\rangle$ and $|l\rangle-|s\rangle\}$ states were prepared on the optical fiber link. Figure 2 shows the temporal intensity profiles of the output from Alice's apparatus and the output from Bob's AMZ, corresponding to the four states prepared. The former was observed through one of the output ports of Alice's AMZ (denoted by Monitor in Fig.1). Since this QKD system relied only on the first-order interference effect, the intensity profiles of the output should be proportional to the probability distributions of photon arrival time when faint laser pulses are used. These distributions agreed well with the expected ones. When we prepared states $\{|l\rangle, |s\rangle\}$, the first and last time slots (denoted by S1 and S3) in the output profiles for Bob differed. However, when we prepared states $\{|l\rangle+|s\rangle, |l\rangle-|s\rangle\}$, the output port for the central slot (denoted by S2) differed. The observed output profiles remained unchanged for more than 30 min. This indicates that our system is useful and sufficiently stable for implementing a practical BB84 QKD system.

In summary, we proposed and constructed an optical interferometer for a BB84

QKD system using PLC-based integrated-optics, and experimentally demonstrated its operation and stability, thereby proving that it is suitable for a BB84 QKD system.

The authors thank Satoshi Ishizaka and Akihisa Tomita for valuable comments. This work was supported by the Telecommunications Advancement Organization of Japan.

**References**

[1]  C. H. Bennett and G. Brassard: in *Proceedings of IEEE International Conference on Computers, Systems, and Signal Processing. Bangalore, India, 1984* (IEEE, New York, 1984), p. 175.

[2]  C. H. Bennett, F. Bessette, G. Brassard, L. Salvail and J. Smolin: J. Cryptol. **5** (1992) 3.

[3]  P. D. Townsend, J. G. Rarity and P. R. Tapster: Electron. Lett. **29** (1993) 634.

[4]  P. D. Townsend, J. G. Rarity and P. R. Tapster: Electron. Lett. **29** (1993) 1291.

[5]  J. D. Franson and B. C. Jacobs: Electron. Lett. **31** (1995) 232.

[6]  C. Marand and P. D. Townsend: Opt. Lett. **20** (1995) 1695.

[7]  A. Muller, H. Zbinden and N. Gisin: Europhys. Lett. **33** (1996) 335.

[8]  A. Muller, T. Herzog, B. Huttner, W. Tittel, H. Zbinden and N. Gisin: Appl. Phys. Lett. **70** (1997) 793.

[9]  H. Zbinden, J. D. Gautier, N. Gisin, B. Huttner, A. Muller and W. Tittel: Electron. Lett. **33** (1997) 586.

[10]  G. Ribordy, J. -D. Gautier, N. Gisin, O. Guinnard and H. Zbinden: Electron. Lett. **34** (1998) 2116.

[11]  H. Kosaka, A. Tomita, Y. Nambu, T. Kimura and K. Nakamura: Electron. Lett. **39** (2003) 1199.

[12]  G. Bonfrate, M. Harlow, C. Ford, G. Maxwell and P. D. Townsend: Electron. Lett. **37** (2001) 846.

[13]  T. Honjo, K. lnoue and H. Takahashi: *Technical Digest of CLEO/IQEC 2004*, (San Francisco, California, May 16-21, 2004), paper ITuJ1.

[14]  T. Kimura, Y. Nambu, T. Hatanaka, A. Tomita, H. Kosaka and K. Nakamura: quant-ph/0403104, 2004, submitted to Jpn. J. Appl. Phys Lett.

[15]  C. H. Bennett: Phys. Rev. Lett. **68** (1992) 3121.

[16]  W. Tittel, J. Brendel, H. Zbinden and N. Gisin: Phys. Rev. Lett. **84** (2000) 4737.

**Figure Captions**

Fig. 1. BB84 QKD system based on integrated-optic interferometers. LS: pulsed light source, SMZ: symmetric Mach-Zehnder interferometer, AMZ: asymmetric Mach-Zehnder interferometer, ATT: attenuator, PM: phase modulator, and APD: avalanche photodiode.

Fig. 2. Temporal intensity profiles of output from Alice's apparatus (upper traces) and output from Bob's AMZ (lower traces) correspond to the four states prepared. In this experiment, we used intense laser pulses. The shaded areas indicate measurements

compatible for discriminating between the two basis states.

Figure 1

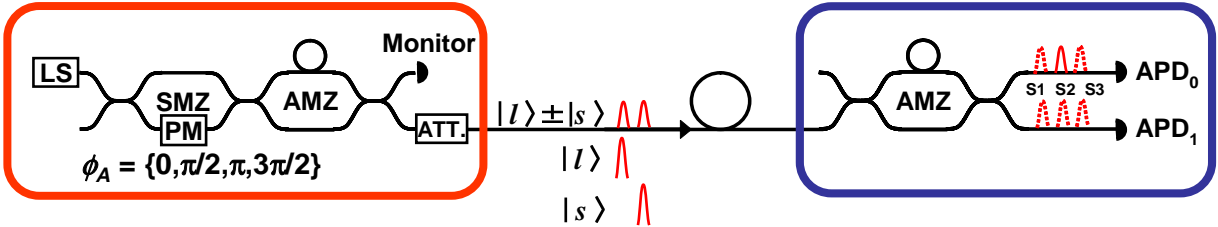

Figure 2

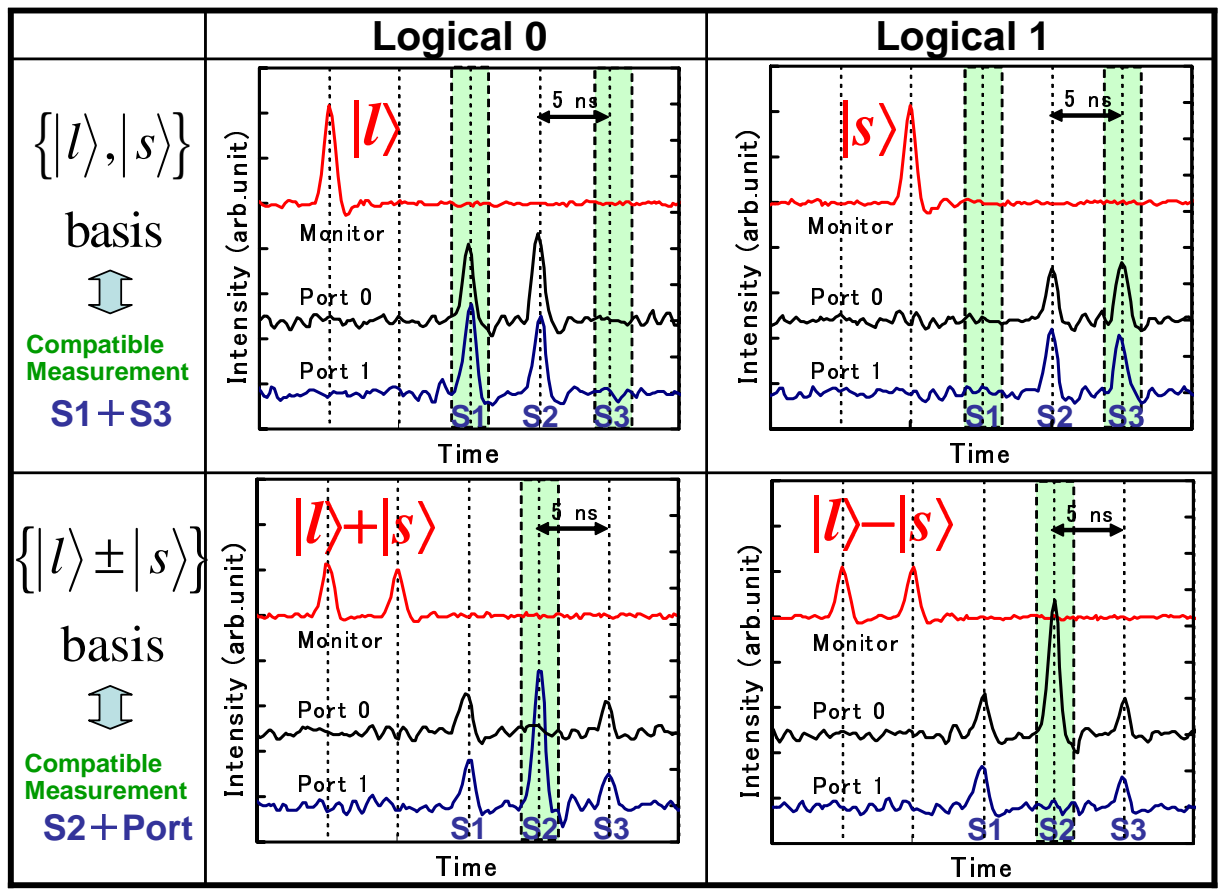